\documentclass[times,twocolumn,final,pdftex]{elsarticle}

\usepackage{medima}
\usepackage[switch,columnwise]{lineno}
\usepackage{amsmath,amssymb,amsfonts}
\usepackage{float}
\usepackage{adjustbox}
\usepackage{stfloats}
\usepackage[justification=justified]{caption}
\usepackage{textcomp}
\usepackage[utf8]{inputenc}
\usepackage{longtable}
\usepackage{url}
\usepackage{xr}
\usepackage{multirow}
\usepackage{xcolor}
\usepackage[colorlinks]{hyperref}
\usepackage{nicefrac}       
\usepackage{microtype}      
\usepackage{booktabs}
\usepackage{multirow}
\usepackage{siunitx}
\usepackage[pdftex]{graphicx}
\usepackage{pifont} 
\newcommand{\cmark}{\ding{51}}%
\newcommand{\xmark}{\ding{55}}%

\hypersetup{
  colorlinks,
  citecolor=teal,
  linkcolor=Blue,
  urlcolor=Blue}

\usepackage{algorithmic}
\usepackage[ruled,vlined]{algorithm2e}
\SetKwInput{KwInput}{Input}                
\SetKwInput{KwOutput}{Output}              
\floatname{algorithm}{Algorithm}

\definecolor{bondiblue}{rgb}{0.0, 0.58, 0.71}
\definecolor{brightcerulean}{rgb}{0.11, 0.62, 0.74}

\def\BibTeX{{\rm B\kern-.05em{\sc i\kern-.025em b}\kern-.08em
    T\kern-.1667em\lower.7ex\hbox{E}\kern-.125emX}}

\journal{Neurocomputing}

\begin{document}
\verso{I. Sivgin, H.A. Bedel \textit{et~al.}}

\begin{frontmatter}
\title{A plug-in graph neural network to boost temporal sensitivity in fMRI analysis}
\author[1,2]{Irmak Sivgin \corref{cor2}}
\author[1,2]{Hasan A. Bedel \corref{cor2}}
\author[1,2,3]{\c{S}aban \"{O}zt\"{u}rk}
\author[1,2,4]{Tolga \c{C}ukur\corref{cor1}}
\cortext[cor2]{Equal contribution}
\cortext[cor1]{Corresponding author, 
  e-mail: cukur@ee.bilkent.edu.tr  }

\address[1]{Department of Electrical and Electronics Engineering, Bilkent University, Ankara 06800, Turkey}
\address[2]{National Magnetic Resonance Research Center (UMRAM), Bilkent University, Ankara 06800, Turkey}
\address[3]{Department of Electrical-Electronics Engineering, Amasya University, Amasya 05100, Turkey}
\address[4]{Neuroscience Program, Bilkent University, Ankara 06800, Turkey}


\begin{abstract}
Learning-based methods have recently enabled performance leaps in analysis of high-dimensional functional MRI (fMRI) time series. Deep learning models that receive as input functional connectivity (FC) features among brain regions have been commonly adopted in the literature. However, many models focus on temporally static FC features across a scan, reducing sensitivity to dynamic features of brain activity. Here, we describe a plug-in graph neural network that can be flexibly integrated into a main learning-based fMRI model to boost its temporal sensitivity. Receiving brain regions as nodes and blood-oxygen-level-dependent (BOLD) signals as node inputs, the proposed GraphCorr method leverages a node embedder module based on a transformer encoder to capture temporally-windowed latent representations of BOLD signals. GraphCorr also leverages a lag filter module to account for delayed interactions across nodes by computing cross-correlation of windowed BOLD signals across a range of time lags. Information captured by the two modules is fused via a message passing algorithm executed on the graph, and enhanced node features are then computed at the output. These enhanced features are used to drive a subsequent learning-based model to analyze fMRI time series with elevated sensitivity. Comprehensive demonstrations on two public datasets indicate improved classification performance and interpretability for several state-of-the-art graphical and convolutional methods that employ GraphCorr-derived feature representations of fMRI time series as their input. 
\end{abstract}


\end{frontmatter}

\section{Introduction}

The human brain comprises networks of regions that interactively process information during cognitive processing \citep{ccukur2013attention}. In turn, correlated activity within individual functional networks has been associated with unique mental states \citep{yan2019reduced, he2018reconfiguration}. Functional MRI (fMRI) is a powerful modality to examine functional networks as it can non-invasively measure whole-brain blood-oxygen-level-dependent (BOLD) signals consequent to neural activity at high spatio-temporal resolution \citep{bai2008default, yuan2008abnormal}. In fMRI studies, functional connectivity (FC) measures are used to assess similarity of BOLD signals among brain regions \citep{cambria2012hourglass, andreu2015big, zhang2015altered, gu2021eeg, yan2018abnormal}. The traditional approach to map FC measures onto mental states is then based on conventional methods such as logistic regression and support vector machines (SVM) \citep{mourao2005classifying, rashid2016classification, dosenbach2010prediction, rosenberg2016neuromarker}. Unfortunately, conventional methods are often insufficiently sensitive to the intricate information patterns in whole-brain fMRI time series \citep{liu2017survey}. 

In recent years, the success of deep learning (DL) models at exploring features in high-dimensional datasets has motivated their adoption for fMRI analysis as an alternative to conventional methods \citep{plis2014deep, kawahara2017brainnetcnn, li2021braingnn, heinsfeld2018identification, kim2016deep}. Earlier attempts in this domain have proposed shallow multi-layer perceptron (MLP) \citep{shen2010discriminative, eslami2019auto} and Boltzmann machine (BM) models \citep{hjelm2014restricted, plis2014deep}. Later studies have adopted deeper architectures based on convolutional neural network (CNN) \citep{zhao2017automatic, kawahara2017brainnetcnn,meszlenyi2017resting}, graph neural network (GNN) \citep{shao2021classification,li2021braingnn, saeidi2022decoding,gadgil2020spatio, kipf2016semi, qu2021brain}, and transformer \citep{yu2022disentangling, malkiel2021pre, nguyen2020attend,dai2022brainformer, bedel2022bolt} models for improved performance. Typically, these models start by constructing construct a set of nodes corresponding to brain regions defined based on an anatomical or functional atlas \citep{tzourio2002automated, rolls2015implementation, laird2009ale}, and receive input features at these nodes based on the FC strength among brain regions \citep{kim2020understanding,li2021braingnn}. A common approach has been to employ static FC features for derived from aggregate correlation measures across the entire duration of fMRI time series \citep{gan2021brain, li2021braingnn}. Yet, this approach is insufficiently sensitive to the dynamic inter-regional interactions in the human brain during resting-state or cognitive tasks \cite{norman2006beyond}. While alternative strategies have recently been proposed to assess the temporal variability in FC features, these methods commonly consider instantaneous signal correlations across local time windows within the time series \citep{kim2021learning, savva2019assessment, handwerker2012periodic}. As such, they do not possess explicit mechanisms to capture delayed correlations between brain regions that can be present in fMRI times series due to hierarchical cognitive processing in the brain or hemodynamic lags in BOLD measurements \citep{celik2021cortical}. 

In this study, we introduce a plug-in graphical neural network, GraphCorr, that provides enhanced input features to learning-based fMRI models so as to boost their sensitivity to dynamic, lagged inter-regional interactions. To capture dynamic changes in interactions, GraphCorr leverages a novel node embedder module based on a transformer encoder that computes hierarchical embeddings of windowed BOLD signals across the time series. To capture lagged interactions between brain regions, GraphCorr employs a novel lag filter module that computes nonlinear features of cross-correlation between pairs of nodes across a range of time delays. The graph model is initialized with node features taken as embeddings from the node embedder module, and with edge weights taken as lag features from the lag filter module. Afterwards, a message passing algorithm is used to compute enhanced node embeddings that account for dynamic, lagged inter-regional interactions. 

Here, we demonstrate GraphCorr for gender classification from fMRI scans in two public datasets: the Human Connectome Project (HCP) dataset \citep{van2013wu} and the ID1000 dataset from Amsterdam Open MRI Collection (AOMIC) \citep{snoek2021amsterdam}. GraphCorr is coupled as a plug-in to state-of-the-art baseline models for fMRI analysis including SAGE \citep{hamilton2017inductive}, BrainGNN \citep{li2021braingnn}, BrainNetCNN \citep{kawahara2017brainnetcnn} and GCN \citep{kipf2016semi}. Significantly enhanced performance is obtained from each baseline model when coupled with GraphCorr. We devise an explanatory analysis approach for GraphCorr to interpret the time frames and brain regions that most significantly contribute to classification decisions. We show that GraphCorr improves explainability of baseline models, resulting in interpretations that are more closely aligned with prominent neuroscientific findings from the literature. We also demonstrate the benefits of GraphCorr-derived features against features extracted via plug-in recurrent neural networks (RNN) and dynamic FC features computed directly from BOLD signals.

\section{Related Work}
Cognitive processes elicit broadly distributed response patterns across the human brain \citep{woolrich2001temporal}. In turn, process-related information such as stimulus or task variables can be decoded by analyzing resultant multi-variate BOLD signals \cite{bolling2011enhanced,shahdloo2020biased}. Initial studies in this domain employed relatively simpler, traditional machine learning (ML) methods for fMRI analysis \citep{norman2006beyond,bu2019investigating}. These traditional methods rely heavily on feature selection procedures to cope with the intrinsically high dimensionality of fMRI data \citep{xie2009brain}. Arguably, FC features among brain regions have been most commonly used to capture discriminative information about cognitive processes \citep{shen2010discriminative, kawahara2017brainnetcnn, kim2020understanding, li2021braingnn}. Many studies have reported that external variables or disease states can be detected given FC features of individual subjects under resting state \citep{wee2012identification, smith2013functional}, cognitive tasks \citep{mourao2005classifying, li2021braingnn}, or both \citep{rosenberg2016neuromarker}. 

Given their earlier success in fMRI analysis, FC features have also been pervasively adopted in recent DL methods that leverage more complex models to enhance performance \citep{riaz2020deepfmri, zeng2018multi}. A pervasive approach in DL-based fMRI analysis relies on static FC features as model inputs, where FC between a pair of regions is taken as the aggregate correlation of their BOLD signals across the entire scan. To extract hierarchical latent representations of these features, earlier studies have proposed either relatively compact fully-connected architectures including BM and MLP models \citep{hjelm2014restricted, plis2014deep, shen2010discriminative, eslami2019auto}, or computation-efficient deep architectures including CNN models \citep{meszlenyi2017resting, kawahara2017brainnetcnn}. Later studies have considered GNN models given their natural fit to analyzing fMRI data that follows an intrinsic connectivity structure \citep{wu2020comprehensive, liu2014distributed, cai2010graph, han2019gcn, li2021braingnn}. These DL methods have all enabled substantial performance improvements in fMRI analysis over traditional methods. Yet, analyses rooted in static FC features can still yield suboptimal sensitivity to fine-grained temporal information across the fMRI time series \citep{rashid2016classification, meszlenyi2017resting}. 

To improve temporal sensitivity, several alternative strategies have been proposed that incorporate time-varying features to DL models for fMRI analysis. A first group of methods pre-compute FC features over moving windows across the time series based on standard correlation measures, and concatenate them across windows to form a higher-dimensional input \citep{sakouglu2010method, allen2014tracking, gadgil2020spatio, savva2019assessment, handwerker2012periodic}. While these dynamic FC features promise enhanced temporal sensitivity, they result in elevated complexity due to their intrinsic dimensionality that can degrade model performance. A second group of methods instead provide voxel-level BOLD signals spatially encoded via a CNN module. Spatially-encoded BOLD signals are then processed with RNN or transformer models to extract the time-varying information \citep{malkiel2021pre, kim2021learning}. Yet, CNN modules based on voxel-level inputs can be difficult to train from scratch under limited data regimes. A third group of methods retain static FC features as their input, albeit augment them with dynamic features captured by RNN modules that directly encode BOLD signals \citep{kim2021learning}. Besides elevated model complexity, these methods can suffer from intrinsic limitations of RNNs in terms of vanishing/exploding gradients over the extensive number of time steps in typical fMRI scans \citep{pascanu2013difficulty, ismail2019input}. Importantly, a common attribute of these previous approaches is that they primarily consider the temporal variation in instantaneous correlations among brain regions. However, they can elicit suboptimal sensitivity as they lack explicit mechanisms to capture delayed inter-regional interactions that can occur due to hierarchical processing or hemodynamic delays \citep{celik2021cortical}. 

Here, we propose to improve the temporal sensitivity of downstream fMRI analysis models by integrating a novel plug-in GNN, GraphCorr. The proposed GraphCorr method uses a novel node embedder module to find contextual embeddings of dynamic FC features based on instantaneous correlations of windowed BOLD signals; and it uses a novel lag filter module to compute embeddings of cross-correlation features of windowed BOLD signals across various time delays. Following a message passing algorithm across the graph, GraphCorr provides enhanced input features that preserve dynamic, delayed correlations among brain regions to a downstream analysis model so as to improve its performance. Unlike methods based on static FC features \citep{kawahara2017brainnetcnn,li2021braingnn}, GraphCorr leverages dynamic FC features to capture the variability in connectivity among brain regions. Unlike methods that receive multiple sets of dynamic FC features across separate time windows \citep{zhang2017hybrid, kim2021learning,gadgil2020spatio}, GraphCorr fuses its node features across time windows to lower model complexity without sacrificing performance. Unlike methods that employ recurrent architectures that involve sequential processing \citep{kim2021learning}, GraphCorr leverages a transformer encoder on dynamic FC features that enables efficient parallel processing. Unlike methods that solely focus on instantaneous signal correlations \citep{kawahara2017brainnetcnn}, GraphCorr adopts an explicit lag filter mechanism to learn delayed cross-correlations among brain regions.

\section{GraphCorr}
\label{corr}

\begin{figure*}[h!]   
  \centering
\includegraphics[width=0.95\linewidth]{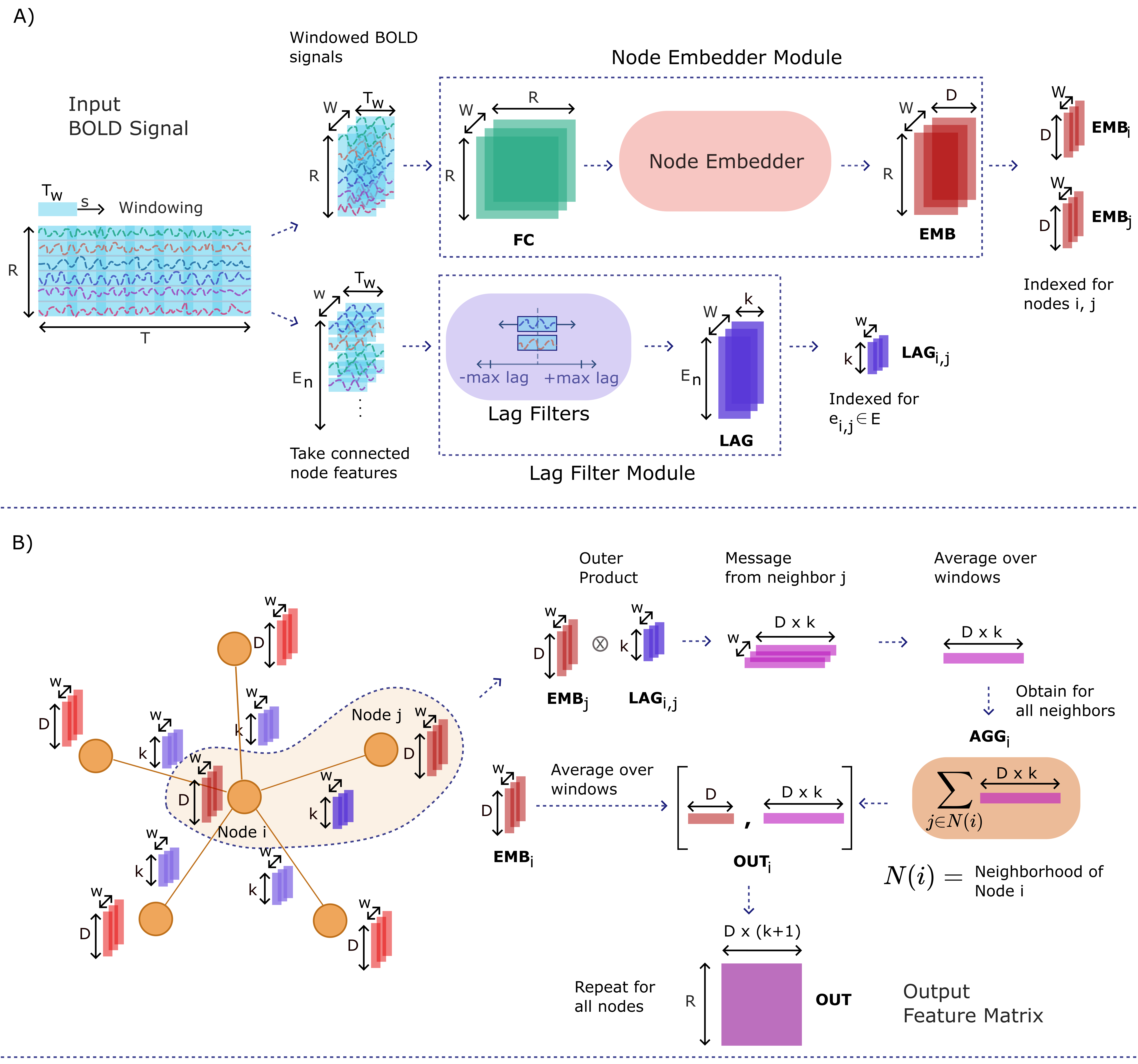}  
  \caption{Overview of GraphCorr.
  \textbf{A.} GraphCorr utilizes two parallel modules to extract dynamic, lagged features of inter-regional correlations across the brain. The node embedder module receives as input time-windowed BOLD signals, and uses a transformer encoder to compute node embeddings of dynamic FC features $\mathbf{EMB} \in \mathbb{R}^{R \times D \times W}$. The lag filter module also receives as input time-windowed BOLD signals, and it computes lag activations due to cross-correlation across a range of lag values $\mathbf{LAG} \in \mathbb{R}^{E_n \times W \times k}$. Cross-correlation is calculated only for connected node pairs ($e_{i,j}=1$). 
  \textbf{B.} To consolidate the extracted feature sets on a graph, node embeddings are taken as node features and lag activations are taken as edge weights. A message passing algorithm is then run on the graph to produce enhanced FC features in an output feature matrix, $\mathbf{OUT} \in \mathbb{R}^{R \times (D \times (k+1))}$.}
  \label{fig:corrfig}
\end{figure*}

Analysis procedures for fMRI time series typically start by defining a collection of $R$ regions of interest (ROI) across the brain based on an anatomical atlas \citep{li2021braingnn, kim2021learning}. Voxel-level BOLD signals within each ROI are then averaged to derive ROI-level signals, resulting in $\mathbf{B} \in \mathbb{R}^{R \times T}$ as the matrix of BOLD signals where $T$ denotes the number of time frames. Static FC features are conventionally computed based on Pearson's correlation coefficient of these BOLD signals across ROIs: $\mathbf{sFC}_{i,j} = \mathrm{Corr}(\mathbf{B}_{i,\cdotp},\mathbf{B}_{j,\cdotp})$, where $\mathbf{sFC} \in \mathbb{R}^{R \times R}$ and $i,j$ are ROI indices. Many previous traditional and learning-based methods use downstream classification models on static FC features, which result in suboptimal temporal sensitivity. Instead, here we propose to extract dynamic FC features of BOLD signals based on a novel GNN plug-in, and to use these enhanced features to improve the performance of downstream classification models. The proposed GraphCorr method forms a graph structure to represent brain connectivity features, leverages node embedder and lag filter modules to capture dynamic, lagged correlations in BOLD signals, and finally performs message passing on the graph to compute enhanced features (Fig. \ref{fig:corrfig}). The methodological components and procedures in GraphCorr are described below.

\subsection{Graph formation}
\label{graph}
As a learning substrate, GraphCorr first forms a graph $G(N,E)$ with $N$ and $E$ denoting nodes and edges, respectively. The node set $N = \{r_i \, | \, i = 1, ..., R\}$ includes ROIs defined according to the atlas, whereas the binary edge set is given as $E = \{e_{i,j} = 1 \, | \, i = 1, ..., R;  j \in \mathcal{N}(i)\}$ where $\mathcal{N}(i)$ is the neighborhood of the $i$-th node. Edges are defined by thresholding to retain the strongest $z\%$ of correlation coefficients in $\mathbf{sFC}$ while excluding self connections, resulting in $E_n$ number of edges. The node features $\mathbf{F} = \{{f}_i \, | \, i = 1, ..., R\}$ are initialized as the time-windowed BOLD signals at each corresponding node to capture local dynamics in the fMRI times series. For this purpose, the scan containing $T$ time frames is split into $W$ windows of size $T_w$ and stride value $s$:
\begin{equation}
    W = \lfloor \frac{T - T_w}{s} \rfloor, 
\end{equation}
resulting in a feature tensor of $\mathbf{F} \in \mathbb{R}^{R \times T_w \times W}$.

\subsection{Architecture}
\label{modules}
To sensitively extract dynamic connectivity features, GraphCorr utilizes a novel node embedding module. To also capture delayed connectivity features, GraphCorr utilizes a novel lag filter module. The two modules are detailed below.

\begin{figure}[h!]   
  \centering
\includegraphics[width=0.7\linewidth]{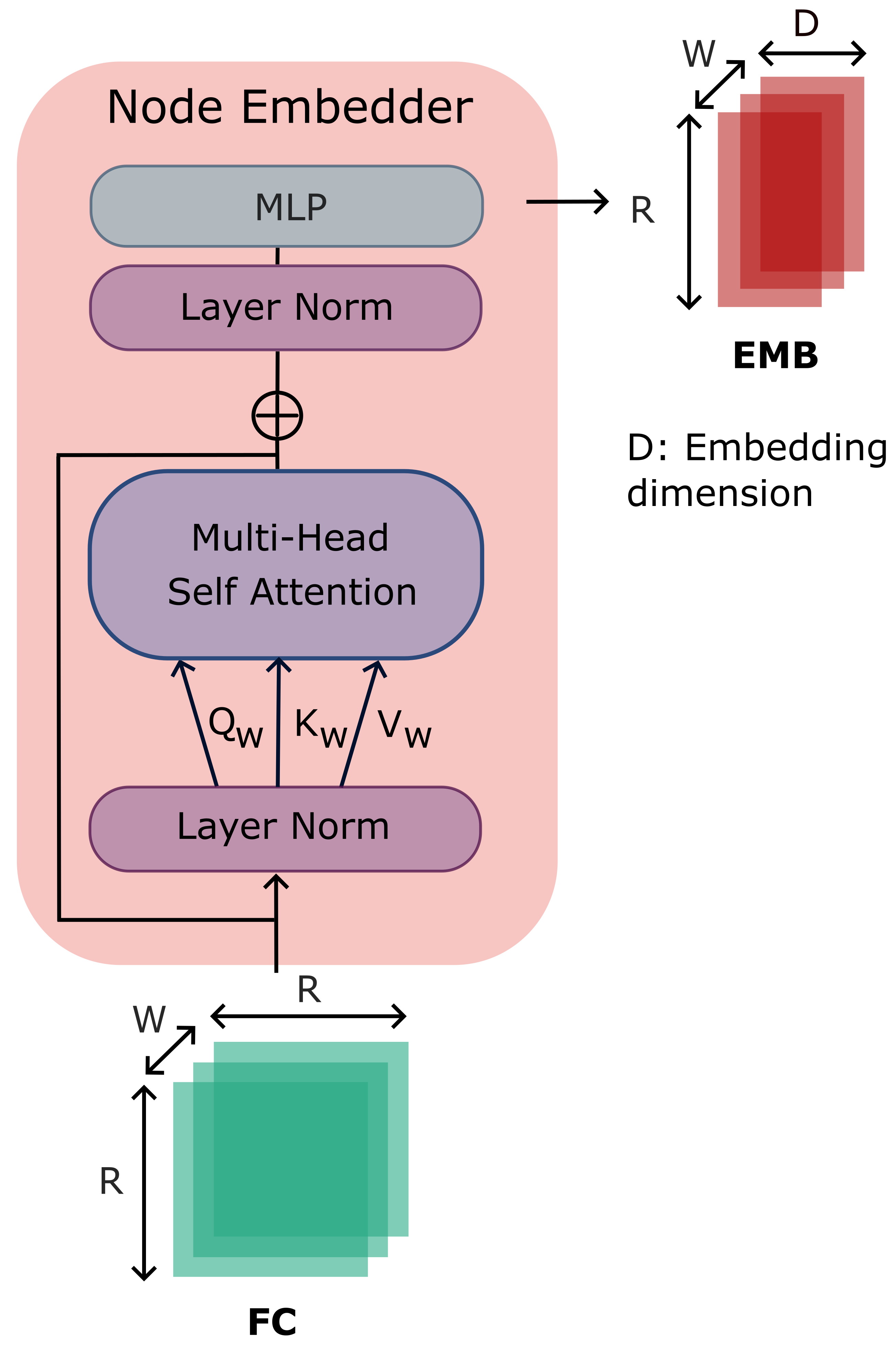}  
  \caption{The node embedder module. The time-windowed FC features $\mathbf{FC}_{\cdotp, \cdotp, w} \in \mathbb{R}^{R \times R}$ at window $w$ are processed with a transformer encoder with multi-head self-attention (MHSA), layer normalization, and multi-layer perceptron (MLP) layers. The output is a node embedding matrix $\mathbf{EMB}_{\cdotp, \cdotp, w} \in \mathbb{R}^{R \times D}$ where $D < R$ denotes the embedding dimensionality.}
  \label{fig:nodeEmbfig}
\end{figure}

\textbf{Node embedder module:} Receiving as input time-windowed BOLD signals, this module computes latent representations of dynamic FC features (Fig. \ref{fig:nodeEmbfig}). First, dynamic FC features are extracted from the time-windowed BOLD signals as: $\mathbf{FC}_{i,j,w} = \mathrm{Corr}(\mathbf{F}_{i,\cdotp,w},\mathbf{F}_{j,\cdotp,w})$ where $w \in \{1,...,W\}$ indicates window index, $i \in \{1,...,R\}$, $j \in \{1,...,R\}$ denote node indices. These FC features are then processed with a transformer encoder where windows across the fMRI time series correspond to the sequence of transformer tokens. Attention calculations are performed on window-specific keys $K_w \in \mathbb{R}^{R \times d}$, queries $Q_w \in \mathbb{R}^{R \times d}$ and values $V_w \in \mathbb{R}^{R \times d}$ derived via learnable linear projections ${f_q, f_k}$ and ${f_v}$: 
\begin{gather}
Q_w = U_q( \{ \mathbf{FC}_{1,\cdotp,w}, \mathbf{FC}_{2,\cdotp,w}, ..., \mathbf{FC}_{R,\cdotp,w} \} ) ,\nonumber\\ 
K_w = U_k( \{ \mathbf{FC}_{1,\cdotp,w}, \mathbf{FC}_{2,\cdotp,w}, ..., \mathbf{FC}_{R,\cdotp,w} \} ) ,\nonumber\\ \label{eqn:qkv}
V_w = U_v( \{ \mathbf{FC}_{1,\cdotp,w}, \mathbf{FC}_{2,\cdotp,w}, ..., \mathbf{FC}_{R,\cdotp,w} \} ),
\end{gather}
where $d$ is the dimensionality of each attention head. The above computations can be performed separately for $H$ attention heads. The window-specific attention matrix $\mathbf{A}_w \in \mathbb{R}^{R \times R}$ is then derived as \citep{dalmaz2022resvit}:
\begin{gather}
\mathbf{A}_w = \mathrm{Att}(Q_w,K_w,V_w) = \mathrm{Softmax}(\frac{Q_w K_w^{\intercal}}{\sqrt{d}}) V_w. \label{eqn:attn}
\end{gather}
Attention matrices are concatenated across attention heads and propagated to an MLP block following layer normalization: 
\begin{gather}
\textbf{EMB}_w = \mathrm{MLP}(\mathbf{A}_w) = \mathrm{GELU}(\mathbf{A}_w \mathbf{M}_{1}) \mathbf{M}_{2} \label{eqn:mlp}
\end{gather}
where $\mathbf{M}_{1} \in \mathbb{R}^{(R \times D)}$ and $\mathbf{M}_{2} \in \mathbb{R}^{(D \times D)}$ denote MLP model parameters, $\mathrm{GELU}$ is the Gaussian activation unit, and $\mathbf{EMB} \in \mathbb{R}^{R \times D \times W}$ are window-specific node embeddings, and $D$ is the embedding dimensionality with $D < R$.

\begin{figure*}[t]   
  \centering
\includegraphics[width=0.9\linewidth]{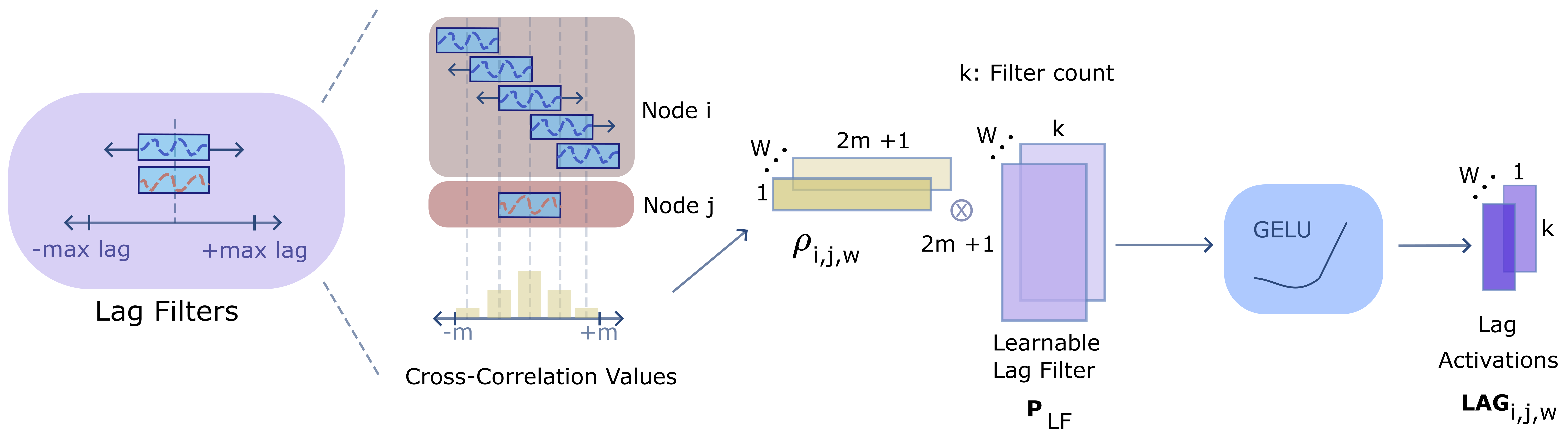}  
  \caption{The lag filter module. Cross-correlation of time-windowed BOLD signals at window $w$ are computed for delays $\tau \in \{ -m, -m+1, ...., m-1, m\}$, where $m$ defines the range. This computation is only performed for pairs of connected nodes ($e_{i,j}=1$). Afterwards, cross-correlation values $\rho_{\cdotp,\cdotp,w,\cdotp} \in \mathbb{R}^{R \times R \times (2m+1)}$ are linearly transformed onto with a learnable filter $M_{LF} \in \mathbb{R}^{(2m+1) \times k}$ onto window-specific lag activations $\textbf{LAG}_{\cdotp,\cdotp,w} \in \mathbb{R}^{R \times R \times k}$.  
}
  \label{fig:lagFilter}
  
\end{figure*}

\textbf{Lag filter module:} Receiving as input time-windowed BOLD signals, this module computes cross-correlation features across a range of temporal delays (Fig. \ref{fig:lagFilter}). For this purpose, initial node features from the graph formation stage are zero-padded across the time dimension: 
\begin{gather}
\mathbf{X}_{i,\cdotp,w} = [\mathbf{0}_{(1\times m)}, \mathbf{F}_{i,\cdotp,w}, \mathbf{0}_{(1\times m)}].
\end{gather}
where $\mathbf{X} \in \mathbb{R}^{R \times (T_w+2m) \times W}$, and $m$ defines the range of delays $\tau \in \{ -m, -m+1, ...., m-1, m\}$ that will be considered in the module. First, cross-correlations are computed between pairs of nodes connected by $e_{ij}$ at each lag value separately: 
\begin{gather}
\rho_{i,j,w,\tau} = \mathrm{Corr} ( \mathbf{X}_{i,\cdotp,w},\mathbf{X}_{j,\cdotp,w},\tau)
\label{eqn:lag}
\end{gather}
Afterwards, learnable lag filters $\mathbf{P}_{LF} \in \mathbb{R}^{(2m+1) \times k}$ with $k$ denoting the number of filters are used to map cross-correlations onto lag activations:
\begin{gather}
\label{eqn:lag2}
\textbf{LAG}_w = \mathrm{GELU}(\rho_{\cdotp,\cdotp,w,\cdotp} \mathbf{P}_{LF})
\end{gather}
where $\mathbf{LAG} \in \mathbb{R}^{E_n \times W \times k}$ are window-specific lag activations.

\subsection{Graph learning} 
\label{message} 
The node embedder produces time-windowed node embeddings, $\textbf{EMB}$, that reflect instantaneous inter-regional correlations. The lag filter produces time-windowed lag activations, $\textbf{LAG}$, that reflect reflect delayed inter-regional correlations. To consolidate these feature sets on the graph, node embeddings are taken as node features and lag activations are taken as edge weights (Fig.   \ref{fig:corrfig}). A message passing algorithm is then run on the graph to compute enhanced FC features. To do this, a message tensor $\textbf{MES} \in \mathbb{R}^{R \times (D\times k) \times W}$ is computed between pairs of connected nodes $(r_i,r_j)$ as:
\begin{gather}
\textbf{MES}_{i,j,\cdotp,w} = {\textbf{EMB}_{j,\cdotp,w}} \textbf{LAG}^{\intercal}_{i,j,\cdotp,w} \label{eqn:msg}
\end{gather}
Messages are first averaged across windows, and then propagated to a target node $r_i$ that sums all messages from one-hop vicinity nodes $j \in \mathcal{N}(i)$:
\begin{gather}
{\textbf{AGG}_i} = \sum_{j\in \mathcal{N}(i)} \frac{1}{W}\sum_{w=1}^{W}{\textbf{MES}_{i,j,\cdotp,w}}. 
\end{gather}
This aggregate message is the concatenated with the window-averaged node embedding at $r_i$:
\begin{gather}
\label{eqn:msg2}
\mathbf{OUT}_i = [\frac{1}{W}\sum_{v=1}^{W}{\textbf{EMB}_{i,\cdotp,w}}, \textbf{AGG}_i ],
\end{gather}
where $\mathbf{OUT}_i \in \mathbb{R}^{D \times (k+1)}$ denotes enhanced features for $r_i$.

\section{Methods}

\subsection{Experimental procedures } \label{Dataset}
Demonstrations were performed on fMRI data from the HCP S1200 release\footnote{\url{https://db.humanconnectome.org}} \citep{van2013wu} and ID1000 dataset from Amsterdam Open MRI Collection (AOMIC)\footnote{\url{https://openneuro.org/datasets/ds003097/versions/1.2.1}} \citep{snoek2021amsterdam}. In the HCP dataset, preprocessed data from resting-state fMRI scans were analyzed. The first resting-state scan among four sessions was selected for each subject, excluding short scans with $T < 1200$. This resulted in a total of 1093 healthy subjects (594 female and 499 male). In the ID1000 dataset, preprocessed data from task-based fMRI scans recorded during movie watching were analyzed. All scans had a fixed duration of $T = 240$. A total of 881 healthy subjects were examined (458 female and 423 male). For both datasets, two alternative ROI definitions were considered, based on either the Schaefer atlas \citep{schaefer2018local} or the AAL atlas \citep{tzourio2002automated}. The Schaefer atlas includes $R = 400$ ROIs within 7 intrinsic networks, whereas the AAL atlas defines $R = 116$ ROIs. 

Experiments were conducted on a single NVIDIA Titan Xp GPU using the PyTorch framework. A nested cross-validation procedure was performed with 5 outer and 1 inner folds  Domain adaptation procedures can be employ to improve reliability  \citep{elmas2022federated}. Data were three-way split into a training set (70\%), a validation set (10\%) and a test set (20\%) with no subject overlap between the sets. For fair comparison, all models were trained, validated and tested on identical data splits. All models were trained based on cross-entropy loss. For each model, hyperparameters were selected to maximize the average performance across the validation sets. A common set of hyperparameters that were observed to yield near-optimal performance were used across datasets and atlases \citep{dalmaz2022one}. Details regarding model implementations are discussed in Section \ref{compare}.

\subsection{Comparative analysis} \label{compare}
GraphCorr was demonstrated on several learning-based methods taken as downstream classification models including SAGE \citep{hamilton2017inductive}, GCN \citep{kipf2016semi}, BrainGNN \citep{li2021braingnn}, and BrainNetCNN \citep{kawahara2017brainnetcnn}. For each method, a vanilla downstream model was trained by providing static FC features as model input, and an augmented downstream model was separately trained where GraphCorr was employed as a plug in to provide model input. Vanilla and augmented models were obtained with identical training procedures. In all graph models, ROIs in a given brain atlas were taken as nodes, and edge selection was then performed based on correlations of BOLD signals. Edges whose correlation coefficients were in the top $z=2\%$ were retained, while remaining edges were discarded. The implementation details of the downstream models and GraphCorr are discussed below. 

\textbf{SAGE:} A GNN model was built based on a module with graph convolution pooling, and fully-connected layers \citep{hamilton2017inductive}. SAGE comprised a cascade of two graphical modules with hidden dimension of 250 and dropout rate of 0.5. Cross-validated hyperparameters were a learning rate of \num{3e-3}, 20 epochs, a batch size of 12. 

\textbf{GCN:} GCN is a GNN model based on graph convolution, pooling and fully-connected layers \citep{kipf2016semi}. GCN comprised a cascade of two graphical modules with hidden dimension of 100 and dropout rate of 0.5. Cross-validated hyperparameters were a learning rate of \num{5e-3}, 30 epochs, a batch size of 12. 

\textbf{BrainGNN:} BrainGNN is a GNN model based on ROI-aware graph convolution, pooling and fully-connected layers \citep{li2021braingnn}. A single graphical module with hidden dimension of 100 and dropout rate of 0.5 was used. Cross-validated hyperparameters were a learning rate of \num{8e-4}, 80 epochs, a batch size of 16.

\textbf{BrainNetCNN:} BrainNetCNN is a CNN model based on convolutional layers with edge-to-edge and edge-to-node filters \citep{kawahara2017brainnetcnn}. The convolutional layers had hidden dimension of 32 and dropout rate of 0.1. Vanilla BrainNetCNN expects a 2D input of size $R \times R$ taken as the static FC matrix. When it was augmented with GraphCorr, its input dimensionality was modified as $R \times D(k+1)$ for compatibility. Cross-validated hyperparameters were a learning rate of \num{2e-4}, 20 epochs, a batch size of 16. 

\textbf{GraphCorr:} The node embedder module was built with a single-layer transformer encoder. Because the scan durations differed across HCP and ID1000, dataset-specific $T_w$ (window size) and $s$ (stride) were selected while common $m$ (maximum lag) and $k$ (filter count) were used. Accordingly, cross-validated parameters were ($T_w$=50, $s$=30, $m$=5, $k$=3) for HCP, ($T_w$=40, $s$=15, $m$=5, $k$=3) for ID1000.

\subsection{Explanatory analysis} \label{interpret}
To assess the influence of GraphCorr on interpretability, the vanilla and augmented versions of trained downstream models were examined. An explanation procedure was devised to identify the brain regions within the fMRI times series that most saliently contribute to the model decisions. First, a gradient-based approach was used to compute a saliency tensor summarizing inter-regional interactions \citep{arslan2018graph, kim2020understanding}. For vanilla models, gradients were computed with respect to static FC features $\mathbf{sFC}$: 
\begin{equation}
\label{eqn:van}
{{\mathbf{SAL}^{van}_{i,j}}} = |\nabla_{\mathbf{sFC}_{i,j}} y_{van}|
\end{equation}
where $\mathbf{SAL}^{van} \in \mathbb{R}^{R \times R}$ and $y_{van}$ denotes the model prediction. For augmented models, gradients were computed with respect to time-windowed FC features ${\mathbf{FC}}_{i,j,w}$: 
\begin{equation}
    \mathbf{SAL}^{aug}_{i,j,w} = |\nabla_{\mathbf{FC}_{i,j,w}} y_{aug}|
\end{equation}
where $\mathbf{SAL}^{aug} \in \mathbb{R}^{R \times R \times W}$ and $y_{aug}$ denotes the model prediction. Afterwards, an ROI-specific saliency score was computed by aggregating values across windows and interacting ROI dimensions of the saliency tensor: 
\begin{gather}
    \mathbf{rSAL}^{van} = \sum_{j=1}^{R}{\mathbf{SAL}^{van}_{\cdotp,j}}\\
    \mathbf{rSAL}^{aug} = \sum_{j=1}^{R} ( \frac{1}{W} \sum_{w=1}^{W}{\mathbf{SAL}^{aug}_{\cdotp,j,w}} )
\end{gather}
where $\mathbf{rSAL}^{van,aug} \in \mathbb{R}^{R}$. For saliency assessment at the level of functional brain networks, the seven intrinsic brain networks defined within the Schaefer atlas were used. For each network, ROI-specific saliency scores were averaged across the regions within the network to obtain a network saliency score per hemisphere.


While unsigned ROI-specific saliency scores reflect the relative importance of each region on the model decision, they do not indicate whether the model output is driven by an increase or decrease in BOLD signals within the ROI. To address this question, a post-hoc logistic regression analysis was conducted. First, important windows in the fMRI time series were determined by aggregating values in the saliency tensor across ROI dimensions: 
\begin{gather}
    \mathbf{wSAL}^{aug} = \sum_{i=1}^{R} \sum_{j=1}^{R}{\mathbf{SAL}^{aug}_{i,j,\cdotp}} \\
    {w}^{*} = \mathop{\arg \max}\limits_{w} {\mathbf{wSAL}^{aug}}
\end{gather}
Here, $ \mathbf{wSAL}^{aug} \in \mathbb{R}^{W}$ denotes the window-specific saliency score used for important window selection. BOLD signals within the most important window were extracted, and thresholded according to intensity to select the top 5 time frames \citep{tagliazucchi2011spontaneous, tagliazucchi2012criticality, bedel2022bolt}. A logistic regression model was then fit to map the BOLD signal vector across ROIs onto the output class, i.e., performing the same task as the downstream model. The logistic model returns a weight for each ROI: a positive weight indicates that an increase whereas a negative weight indicates that a decrease in the ROI's BOLD signal elicits the downstream model's decision.


\begin{table}
  \caption{Performance of downstream models on the HCP and ID1000 datasets with the Schaefer atlas. Results are listed as mean$\pm$std across test folds for vanilla and GraphCorr-augmented versions. Boldface indicates the better performing version of each model.}
  \label{SchaeferResults}
  \centering
  \resizebox{\columnwidth}{!}{
  \begin{tabular}{ll ll ll}
    \toprule
    \multirow{2}{*}{Model} & {}&\multicolumn{2}{c}{HCP} & \multicolumn{2}{c}{ID1000}  \\
    \cmidrule(lr){3-4} \cmidrule(lr){5-6}     
    
    & {}&
            
    Acc (\%)   & ROC (\%) &
    
    Acc (\%)  & ROC (\%) \\
    
    \midrule
    \multirow{2}{*}{SAGE}& Vanilla & $75.2 \pm 2.84 $ & $85.29 \pm 1.67$ & $62.39 \pm 2.17$ & $68.59 \pm 3.76$ \\
    {}& Augmented & $\textbf{89.57} \pm 0.68$  & $\textbf{94.27} \pm 1.98$ & $\textbf{81.7} \pm 2.67$ & $\textbf{87.02} \pm 2.10$   \\
    \midrule
    \multirow{2}{*}{GCN} & Vanilla & $79.14 \pm 2.93$   & $86.00 \pm 1.47$ & $67.84  \pm 2.95$ & $71.78 \pm 3.88$   \\
    {}& Augmented  & $\textbf{89.94} \pm 2.18$  & $\textbf{94.52} \pm 1.61$ & $\textbf{80.80} \pm 0.98$  & $\textbf{87.90} \pm 1.75$   \\
    \midrule
    \multirow{2}{*}{BrainGNN}  & Vanilla & $72.83 \pm 1.98$ & $78.85 \pm 2.45$  & $62.50 \pm 1.80$ & $65.63 \pm 2.81$ \\
    {}& Augmented  & $\textbf{84.72} \pm 1.33$  & $\textbf{92.97} \pm 0.98$ & $\textbf{79.32} \pm 1.96$ & $\textbf{88.10} \pm 2.19$   \\
    \midrule
    \multirow{2}{*}{BrainNetCNN}  & Vanilla & $82.52 \pm 2.80$   & $91.23 \pm 1.21$ & $ 75.45 \pm 2.01$ & $83.65 \pm 2.05$   \\
    {}& Augmented  & $\textbf{88.47} \pm 2.63$  & $\textbf{94.71} \pm 1.86$ & $\textbf{82.73} \pm 1.63$  & $\textbf{89.85} \pm 2.24$   \\
    
    \bottomrule
  \end{tabular}}
\end{table}
\begin{table}
  \caption{Performance of downstream models on the HCP and ID1000 datasets with the AAL atlas. Results are listed as mean$\pm$std across test folds for vanilla and GraphCorr-augmented versions. Boldface indicates the better performing version of each model.}
  \label{AALResultTable}
  \centering
  \resizebox{\columnwidth}{!}{
  \begin{tabular}{ll ll ll}
    \toprule
   \multirow{2}{*}{Model} & {}&\multicolumn{2}{c}{HCP} & \multicolumn{2}{c}{ID1000}  \\
    \cmidrule(lr){3-4} \cmidrule(lr){5-6} 
    
    & {}&
            
    Acc (\%)   & ROC (\%) &
    
    Acc (\%)  & ROC (\%) \\
    
    \midrule
    \multirow{2}{*}{SAGE}& Vanilla & $68.26 \pm 3.31 $ & $75.65 \pm 1.29$  & $62.84 \pm 1.81$ & $67.23 \pm 2.71$ \\
    {}& Augmented & $\textbf{85.45} \pm 3.57$  & $\textbf{91.19} \pm 2.63$ & $\textbf{77.50} \pm 3.68$ & $\textbf{84.92} \pm 1.92$   \\
    \midrule
    \multirow{2}{*}{GCN} & Vanilla & $69.90 \pm 1.35$   & $75.85 \pm 1.06$ & $65.45  \pm 1.36$ & $70.96 \pm 1.23$   \\
    {}& Augmented  & $\textbf{84.36} \pm 3.17$  & $\textbf{88.99} \pm 2.83$ & $\textbf{79.43} \pm 3.45$ & $\textbf{85.43} \pm 2.49$   \\
    \midrule
    \multirow{2}{*}{BrainGNN}  & Vanilla  & $65.69 \pm 3.00$ & $71.79 \pm 2.97$  & $62.27 \pm 3.44$ & $66.42 \pm 4.24$ \\
    {}& Augmented  & $\textbf{80.60} \pm 2.67$  & $\textbf{89.32} \pm 2.67$ & $\textbf{75.11} \pm 0.56$ & $\textbf{83.14} \pm 1.41$   \\
    \midrule
    \multirow{2}{*}{BrainNetCNN}  & Vanilla  & $68.16 \pm 3.53$   & $74.76 \pm 2.08$ & $ 75.00 \pm 2.19$ & $81.44 \pm 3.27$   \\
    {}& Augmented  & $\textbf{83.99} \pm 2.92$  & $\textbf{91.25} \pm 2.65$ & $\textbf{78.98} \pm 1.83$  & $\textbf{86.16} \pm 0.84$   \\

    \bottomrule
  \end{tabular}}
\end{table}

\section{Results}
 \label{results}
\subsection{Comparative analysis}
GraphCorr was demonstrated on downstream classification models based on SAGE \citep{hamilton2017inductive}, GCN \citep{kipf2016semi}, BrainGNN \citep{li2021braingnn}, and BrainNetCNN \citep{kawahara2017brainnetcnn}. A gender detection task was performed given resting-state fMRI scans in individual subjects. Performances of vanilla and augmented versions of downstream models on HCP and ID1000 datasets are listed in Table \ref{SchaeferResults} for the Schaefer atlas, and in Table \ref{AALResultTable} for the AAL atlas. In all examined cases, augmentation with GraphCorr significantly enhances the performance of downstream models (p$<$0.05, Wilcoxon signed-rank test). When ROIs are defined via the Schaefer atlas, GraphCorr enables (accuracy,\,ROC)\% improvements of (14.37,\,8.98)\% for SAGE, (10.80,\,8.52)\% for GCN, (11.89,\,14.12)\% for BrainGNN, and (5.95,\,3.48)\% for BrainNetCNN on HCP; and it enables improvements of (19.31,\,18.43)\% for SAGE, (13.32,\,16.12)\% for GCN, (16.82,\,22.47)\% for BrainGNN, and (7.28,\,6.20)\% for BrainNetCNN on ID 1000. When ROIs are defined via the AAL atlas, GraphCorr enables improvements of (17.19,\,15.54)\% for SAGE, (14.46,\,13.14)\% for GCN, (14.91,\,17.53)\% for BrainGNN, and (15.83,\,16.49)\% for BrainNetCNN on HCP; and it enable improvements of (14.66,\,17.69)\% for SAGE, (13.98,\,14.47)\% for GCN, (12.84,\,16.72)\% for BrainGNN, and (3.98,\,4.72)\% for BrainNetCNN on ID1000. We observe that for vanilla versions of the relatively simpler GNN models perform poorly against the more complex BrainNetCNN model. However, GraphCorr-augmented versions of these GNN models start outperforming the augmented BrainNetCNN. Thus, our results suggest that the feature extraction capabilities of vanilla GNN models might be suboptimal in comparison to CNN-based architectures, albeit a powerful feature extractor on the input side can mitigate this deficit in favor of GNN models.

\subsection{Explanatory analysis}
To assess the influence of GraphCorr on interpretability, an explanatory analysis was conducted separately on the trained vanilla and augmented downstream models. For this analysis, the HCP dataset and the Schaefer atlas were selected that have been broadly studied in the literature for intrinsic brain networks during resting state \citep{smith2013resting}. First, network saliency scores obtained in each hemisphere were compared between vanilla and augmented versions of SAGE, which generally maintains the highest performance after GraphCorr augmentation. Literature reports that BOLD signals across the sensorimotor network (SMN), the default mode network (DMN) and the visual network bilaterally across the two hemispheres carry discriminative information on subject gender \citep{zhang2018functional,kim2020understanding}. Accordingly, we reasoned that a successful downstream classification model for gender detection should focus on these networks. Vanilla SAGE shows somewhat heterogeneous results with significant salience in the DMN in the left hemipshere (LH); the attention network, the SMN, and the visual network in the right hemisphere (RH); but unexpectedly it also yields strong salience in the limbic network in both hemispheres ($p<0.05$, Wilcoxon signed-rank test) although it is not considered to carry information on gender. In contrast, augmented SAGE shows significant salience across the SMN, the DMN and the attention network in the RH; the visual network in both hemipsheres ($p<0.05$), without any salience in the limbic network. These results imply that GraphCorr helps improve interpretability of the downstream classification model by allowing it to focus on brain regions that carry task-relevant information.

\begin{figure}[h!]   
  \centering
\includegraphics[width=0.85\linewidth]{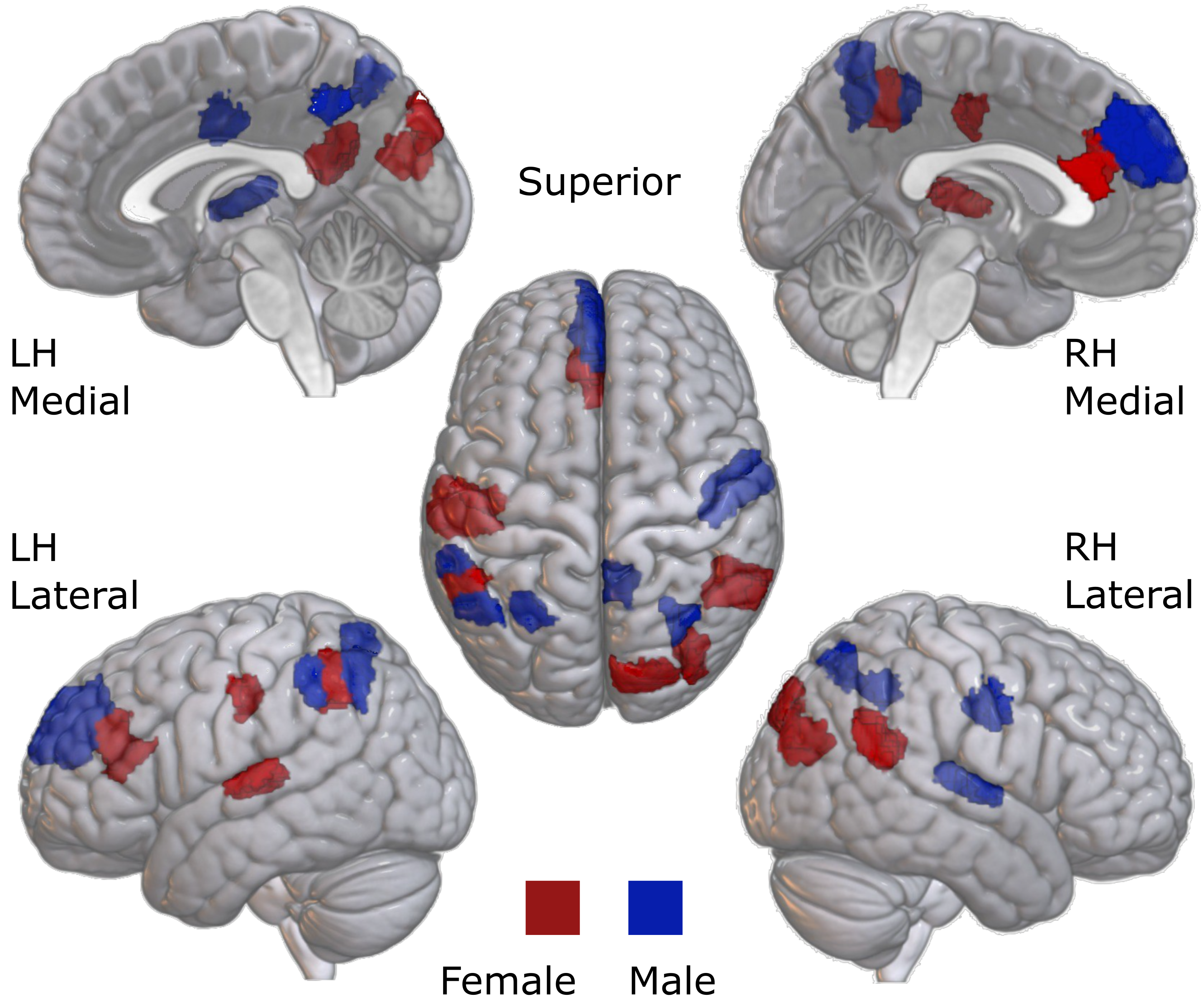}  
  \caption{Salient ROIs for gender detection assessed via the logistic regression analysis. Results are shown for the GraphCorr-augmented SAGE model on the HCP dataset with the Schaefer atlas. ROIs with the top 2\% saliency scores are marked. Red color indicates ROIs whose BOLD signals are elevated in female subjects, whereas blue color indicates ROIs whose BOLD signals are elevated in male subjects.}
  \label{fig:brain}
\end{figure}

Significant saliency in a network indicates that BOLD signals across that network carry information about subject gender. Yet, it does not explain whether an increase or decrease in BOLD signals is evoked for individual genders. To address this question, a logistic regression analysis was conducted on BOLD signals extracted from the most important time window determined according to ROI saliency scores. Specifically, a logistic regression model was fit to detect subject gender given important BOLD signals. Fig. \ref{fig:brain} illustrates the ROI weights in the logistic model, where a positive weight indicates that elevated BOLD signals in the ROI are associated with female subjects, and a negative weight indicates that elevated BOLD signals in the ROI are associated with male subjects. Accordingly, ROI weights were inspected for the logistic regression analyses based on the augmented SAGE model. In females, elevated BOLD signals are identified in RH parietal DMN areas including posterior cingulate cortex (PCC), RH prefrontal DMN areas, LH prefrontal control areas and LH-RH SMN areas. In males, elevated BOLD signals are identified in LH prefrontal DMN areas, LH-RH dorsal attention areas (DAN), LH parietal and prefrontal control areas, and RH parietal and extrastriate visual areas. These findings are consistent with evidence that females have relatively higher activations across DMN areas including PCC, and that males have relatively higher activations in visual and attentional areas \citep{ritchie2018sex,allen2011baseline}. Our results are also consistent with recent studies suggesting that SMN and prefrontal regions show discriminate activation patterns across the two genders \citep{kim2020understanding}. 

\subsection{Ablation studies}
Ablation studies were performed to assess the contribution of the individual design elements in GraphCorr to model performance. These analyses were conducted based on the SAGE model using the HCP dataset and the Schaefer atlas, i.e., the setting that yields the highest overall performance for gender detection. First, we assessed contributions of the node embedder module, lag filter module, and time windowing in GraphCorr. To ablate the node embedder module, node embeddings prior to message passing were initialized with the unlearned time-windowed FC matrix derived via conventional correlation measures on BOLD signals. To ablated the lag filter module, a single filter at zero lag was used within the module to consider only instantaneous correlations. To ablated time windowing, the entire fMRI time series was provided to GraphCorr with a single window of size equal to scan duration. Table \ref{ablationSchaefer} lists performance metrics for ablated variants of GraphCorr. We find that the node embedder module, the lag filter module and time windowing enable (accuracy,\, ROC)\% improvements of (5.31,\,3.2)\%, (0.65,\,0.17)\%, and (8.41,\,5.61)\%, respectively. 

Next, we assessed the benefits of GraphCorr over alternative plug-in approaches to improve the temporal sensitivity of downstream model. In particular, we considered providing the downstream model pre-computed dynamic FC features across time windows via conventional correlation measures \citep{zhang2017hybrid}, an RNN model based on LSTM layers \citep{dvornek2017identifying}, and an RNN model based on GRU layers \citep{gao2022age}. The feature dimensionality at the output of all plug-in models were identical to that for GraphCorr. Table \ref{alternativeRNN} lists performance metrics for different plug-in methods. GraphCorr outperforms all other plug-in methods, with (5.77,\,2.15)\% higher performance than the top-contending GRU method.

\begin{table}[t]
  \caption{Performance for ablated variants of GraphCorr on the HCP dataset with the Schaefer atlas. Results are listed as mean$\pm$std across test folds for the downstream SAGE model. Boldface indicates top-performing variant.}
  \label{ablationSchaefer}
  \centering
    \resizebox{0.8\columnwidth}{!}{
  \begin{tabular}{p{1.5cm} p{1.5cm} p{1.5cm} l l p{2cm} p{2cm}}
    \toprule
    \centering Node Embedder & \centering Lag Filter & \centering Windowing & {Accuracy (\%)} & {ROC (\%)} \\    
    
    \midrule
     \centering \xmark & \centering \xmark  & \centering \xmark  & $75.20 \pm 2.83$  & $85.29 \pm 1.67$ \\   
     \centering \cmark & \centering \xmark  & \centering \xmark  & $80.51 \pm 1.92$ & $88.49 \pm 1.54$ \\
     \centering \cmark & \centering \cmark  & \centering \xmark  & $81.16 \pm 1.69$ & $88.66 \pm 2.07$ \\
     \centering \cmark & \centering \cmark  & \centering \cmark  & $\textbf{89.57} \pm 0.68$ & $\textbf{94.27} \pm 1.98$ \\

    \bottomrule
  \end{tabular}}
\end{table}

    


\begin{table}[t]
  \caption{Performance of competing plug-in methods on the HCP dataset with the Schaefer atlas. Results are listed as mean$\pm$std across test folds for the downstream SAGE model. Boldface indicates top-performing plug-in.}
  \label{alternativeRNN}
  \centering
    \resizebox{0.55\columnwidth}{!}{
  \begin{tabular}{ccc}
    \toprule
     \multirow{1}{*}{Plug-in} & \multirow{1}{*}{Accuracy (\%)}  & \multirow{1}{*}{ROC (\%)}  \\

    \midrule
    \centering {Dynamic FC} & $81.06 \pm 2.94$  & $89.04 \pm 1.53$ \\   
    \centering {LSTM} & $83.26 \pm 2.05$  & $90.29 \pm 1.40$ \\   
    \centering {GRU} & $83.80 \pm 2.29$  & $92.12 \pm 2.01$ \\
    \centering {GraphCorr} & $\textbf{89.57} \pm 0.68$  & $\textbf{94.27} \pm 1.98$ \\   
    \bottomrule
  \end{tabular}}
\end{table}

\section{Discussion}


Here we reported a novel plug-in GNN method, GraphCorr, to improve the performance of downstream classification models in fMRI analysis by capturing dynamic, lagged FC features of BOLD signals. Demonstrations were provided on two large-scale resting-state fMRI datasets, where substantially improved performance was achieved following model augmentation with GraphCorr. The proposed method can be trivially combined with classification models to detect other categorical variables related to cognitive task or disease \cite{li2021braingnn, kawahara2017brainnetcnn}. Alternatively, it can be employed as a plug-in to downstream regression models to boost sensitivity in predicting continuous variables related to stimulus or task features \citep{ccukur2013attention}.  

A mainstream approach in neuroimaging studies rests on prediction of experimental variables typically related to stimulus or task from BOLD signals \citep{norman2006beyond,shahdloo2020biased}. Here we adopted this approach to build decoding models that predict subject gender from resting-state fMRI scans. An alternative procedure to examine cortical function rests on encoding models that instead predict BOLD signals from experimental variables \citep{celik2021cortical,shahdloo2022task, anderson2016representational}. It may be possible to adopt GraphCorr to improve sensitivity of such downstream encoding models. In this case, GraphCorr would receive as input the time course of experimental variables during an fMRI scan. In turn, it would learn dynamic, lagged correlations among experimental variables to better account for their distribution. Learned correlations might help improve performance of downstream regression models that aim to predict measured BOLD signals. Future work is warranted to investigate the potential of GraphCorr in building encoding models for fMRI. 

In conjunction with downstream models, GraphCorr was directly trained end-to-end on the HCP or ID1000 datasets that contained data from several hundred subjects. While the lag filter module has low complexity, the node embedder module uses a transformer encoder with a relatively large number of parameters. To improve learning on limited datasets, transfer learning can be performed where the encoder is initialized with pre-trained weights \citep{korkmaz2022unsupervised}. Data augmentation procedures that can produce a large variety of realistic samples from a learned distribution might further facilitate learning \citep{dar2022adaptive,ozbey2022unsupervised}. GraphCorr forms an initial graph where edges are retained in a single-hop neighborhood based on static FC values between corresponding nodes. This structure is kept fixed during subsequent training procedures. To improve performance, an adaptive structure can be used instead where the edge weights are taken as learnable parameters.

Here each individual subject's fMRI scans were aligned to an anatomical template, and brain regions were then defined with guidance from a brain atlas. The mean BOLD signals in each ROI were then processed in downstream models. Benefits of this approach include computational efficiency due to relatively lower model complexity, and consistency in region definitions across subjects \citep{flandin2002improved}. Meanwhile, information losses naturally occur during registration of individual-subject fMRI data onto a standardized template. To alleviate these losses, ROI definitions in the template space could instead be backprojected onto the brain spaces of individual subjects. This way ROI definitions can be performed while leaving fMRI data in its original space \citep{shahdloo2020biased}.

\section{Conclusion}
In this study, we introduced a novel plug-in graph neural network to improve the performance of downstream models for fMRI classification. The proposed GraphCorr method employs node embedder and lag filter modules to sensitively extract dynamic and lagged functional connectivity features from whole-brain fMRI time series. As such, it transforms raw BOLD signals into a graph representation where neighboring nodes are taken as brain regions with correlated signals and node features are extracted via message passing on connectivity features from the two modules. This procedure restores the fine-grained temporal information that can otherwise be diminished in conventional functional connectivity features. As augmenting downstream classification models with GraphCorr significantly improves their performance and interpretability, GraphCorr holds great promise for analysis of fMRI time series.

\section*{Acknowledgments}
This study was supported in part by a TUBITAK BIDEB scholarship awarded to H.A. Bedel, by TUBA GEBIP 2015 fellowship, BAGEP 2017 fellowship, and TUBITAK 121N029 grant awarded to T. Çukur.

\bibliographystyle{model3-num-names}
\bibliography{Papers}

\end{document}